\documentclass[aps,reprint, twocolumn,
superscriptaddress,amsmath,amssymb]{revtex4-1}%
\usepackage{epsf}
\usepackage{graphicx}
\usepackage{color}
\usepackage{amsmath}
\usepackage{amsfonts}
\usepackage{amssymb}
\usepackage{dcolumn}
\usepackage{txfonts}%
\providecommand{\U}[1]{\protect\rule{.1in}{.1in}}

\begin{document}

\title{Magnetic-proximity-induced magnetoresistance on topological insulators}
\date{\today}
\date{\today}
\author{Takahiro Chiba}
\email{t.chiba@imr.tohoku.ac.jp}
\affiliation{Institute for Materials Research, Tohoku University, Sendai 980-8577, Japan}
\author{Saburo Takahashi}
\affiliation{Institute for Materials Research, Tohoku University, Sendai 980-8577, Japan}
\author{Gerrit E. W. Bauer}
\affiliation{Institute for Materials Research, Tohoku University, Sendai 980-8577, Japan}
\affiliation{WPI-AIMR, Tohoku University, Sendai 980-8577, Japan}
\affiliation{Kavli Institute of NanoScience, Delft University of Technology, Lorentzweg 1,
2628 CJ Delft, The Netherlands}

\begin{abstract}
We theoretically study the magnetoresistance (MR) of two-dimensional
massless Dirac electrons as found on the surface of three-dimensional
topological insulators (3D TIs) that is capped by a ferromagnetic
insulator (FI). We calculate charge and spin transport by Kubo and
Boltzmann theories, taking into account the ladder-vertex correction
and the in-scattering due to normal and magnetic disorder. The induced
exchange splitting is found to generate an electric conductivity that
depends on the magnetization orientation, but its form is very different
from both the anisotropic and spin Hall MR. The in-plane MR vanishes
identically for nonmagnetic disorder, while out-of-plane magnetizations
cause a large MR ratio. On the other hand, we do find an in-plane
MR and planar Hall effect in the presence of magnetic disorder aligned
with the FI magnetization. Our results may help understand recent
transport measurements on TI
$\vert$
FI systems.
\end{abstract}
\maketitle


\section{Introduction}

The control of electric transport by utilizing the spin angular momentum
has been a central theme in spintronics after the discovery of the
giant and tunnel magnetoresistances, leading to new functionalities
for sensing, logic, and data storage applications \cite{Hoffmann15}.
On the other hand, the anisotropic magnetoresistance (AMR) in ferromagnets,
i.e. the dependence of electric transport on the relative angle between
the current and magnetization directions, has been already discovered
in 1857 by Lord Kelvin \cite{AMR}. Just like the anomalous Hall effect
(AHE), it is rooted in the spin-orbit coupling (SOC). In the absence
of a general theory, several studies addressed the AMR in a simple
model system, viz. the two-dimensional (2D) electron gas with Rashba
and Dresselhaus SOCs. The applied methods were the Boltzmann equation
\cite{Schliemann03,Vyborny09} and the linear-response Kubo formalism
\cite{Kato08,Kovalev09}.

Recently, a so-called spin Hall magnetoresistance (SMR) has been discovered
in bilayers made from heavy normal metals such as platinum and ferromagnetic
insulators (FIs) such as $\mathrm{Y_{3}Fe_{5}O_{12}}$ (YIG) \cite{Nakayama13,Chen13}.
The SMR can be explained by the simultaneous action of the spin Hall
effect (SHE) \cite{Sinova14} and its inverse that is modulated by
the spin transfer torque or relative angle of the current-induced
spin polarization in the metal and the magnetization direction of
the ferromagnet. Hence, the SMR\ is a nonlocal and nonequilibrium
magnetic proximity effect (MPE). Alternative mechanisms for the SMR
have been proposed, i.e., the magnetized normal metal, typically Pt,
by the ferromagnet contact \cite{Miao14} or the Rashba SOC at the
interface \cite{Grigoryan14,Zhang15}, but theoretical and experimental
support of those models is still scarce. SMR-like phenomena have been
observed for all metallic bilayers as well \cite{Avci15,Kim16}, but
the interpretation of the results is easier when the magnet is an
electric insulator since parallel current paths through the magnet
can be excluded. The reported SMR ratios are quite small (of the order
of $10^{-4}$), being proportional to the squared spin Hall angle
which is typically less than 10\% \cite{Sinova14}.

Three-dimensional topological insulators (3D TIs) are ideally insulating
in the bulk while supporting topologically protected metallic surface
states as a consequence of time reversal symmetry and band inversion
induced by a strong SOC \cite{Hasan10,Qi11,Ando13}. In the surface
of TIs as well as the Rashba-splitting 2D electron gas (2DEG) the
helical band structure is realized, in which the spin and momentum
are locked and hence the surface currents are spin-polarized \cite{Li14,Shiomi14,Kondou15}.
The interface between a TI (or a Rashba 2DEG \cite{Burkov04}) and
a ferromagnet can be a spin source in which the SOC enhances the magnitude
of both charge and spin currents \cite{Yokoyama10,Burkov10,Jiang14,AndoI14}.
Electric transport properties of bilayers of 3D TIs with (metallic)
ferromagnets have indeed been interpreted in terms of much larger
spin Hall angles \cite{Fan14,Mellnik14}. Recently, there have been
some experiments with YIG for the spin-charge conversion \cite{Jiang16,Wang16}.
TIs are therefore a promising platform to enhance the SMR.

The SMR interpretation in terms of SHE and inverse SHE is based on
semiclassical spin diffusion model and does not hold for 2D materials.
Since the transport is confined now to an atomic monolayer, an MR
generated by an induced proximity exchange potential (or equilibrium
MPE) appears plausible \cite{Jiang15,Wang15,Leutenantsmeyer16,Wei16}.
In 3D-systems this effective interface magnetic field is proportional
to the imaginary part of the mixing conductance \cite{Tserkovnyak05}
that for an interface between a FI and a nonmagnetic metal is relatively
small and is usually disregarded \cite{Jia11}. For graphene on YIG,
a proximity potential of 20\,$\mathrm{\mu eV}$\ (0.2\,T) has been
reported \cite{Leutenantsmeyer16}, which is smaller than predicted
\cite{Hallal16}. A much larger proximity potential of $14\,\mathrm{meV}$
(14 T) has been reported for graphene on EuS \cite{Wei16}.

In spite of the progress in understanding the magnetoresistance (MR)
of a magnetized 2DEG with Rashba SOC and the large attention for the
AHE in Zeeman-split TI surface states, a thorough discussion of the
AMR/SMR\ of the latter appears to be lacking. We therefore report
here a theory of the MR of a TI$|$FI bilayer, modeled as a 2D Dirac
system with finite exchange splitting, where the latter is a vector
parallel to the FI magnetization that can be controlled by applied
magnetic fields \cite{Wang15,Wei16}. We calculate the electric dc
conductivity in magnetized 2D Dirac electron system by the Kubo formalism
and the linearized Boltzmann equation with random potential disorder.
The equilibrium magnetic proximity effect, i.e. the exchange interaction
in the surface state induced by an attached magnet, is found to generate
an MR that depends on the magnetization orientation. However, its
form differs from both AMR and SMR. For in-plane magnetizations the
MR vanishes identically in the TI$|$FI bilayer, while an out-of-plane
magnetization causes a large MR ratio. Moreover, we do find an in-plane
MR and planar Hall effect in the presence of magnetic disorder when
aligned with the FI magnetization. Our calculated results agree well
with the MR observations. We also discuss the current-induced spin
polarization and the role of magnetic impurities.

In Sec.~\ref{2D Dirac model}, we present a model for the surface
of TIs with a finite exchange potential controlled by an FI contact.
In Sec.~\ref{Kubo}, we calculate the electric dc conductivity in
magnetized 2D Dirac electrons with randomly distributed nonmagnetic
disorder by the Kubo formalism. In Sec.~\ref{Boltzmann}, we address
the same problem by the linearized Boltzmann equation and get identical
results. We also discuss the current-induced spin polarization. In
Sec.~\ref{magnetic impurities}, we address the effect of magnetic
impurities on electric transport and briefly discuss the related MR
experiments on TI. We summarize the results and conclusions in Sec.~\ref{Summary}.


\section{Two-dimensional massless Dirac model}

\label{2D Dirac model} 
\begin{figure}[ptb]
\begin{centering}
\includegraphics[width=0.3\textwidth,angle=0]{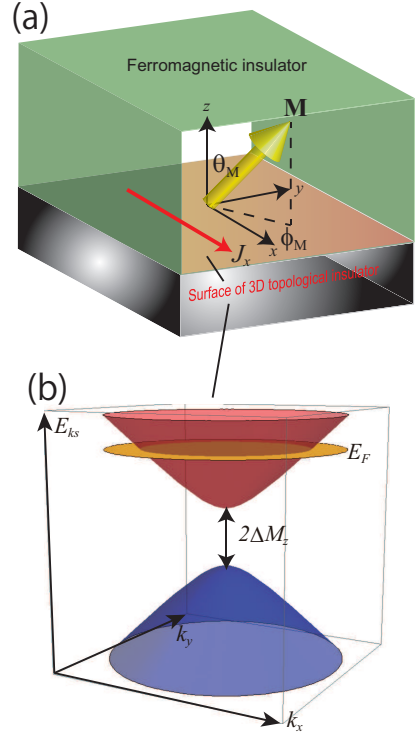} 
\par\end{centering}
\caption{(a) Bilayer of a three-dimensional topological insulator and a ferromagnetic
insulator. Electric currents flow on the surface of the TI in proximity
of the magnet (shown as the red area). (b) Schematic energy dispersion
of the gapped 2D Dirac Hamiltonian. The Fermi level is taken to be
above the gap. }
\label{fig:system2d}
\end{figure}

We consider 2D massless Dirac electrons on
the surface of the TI, exchange-coupled to a homogeneous magnetization
of an attached FI, as shown in Fig.~\ref{fig:system2d}.
A simple model for the electronic structure of a TI surface state
is the massless Dirac Hamiltonian \cite{Hasan10}. When the TI electrons
are in contact with an FI \cite{Jiang15}, they experience an exchange
interaction that can be modeled by a constant spin splitting $\Delta$
along the magnetization direction with unit vector $\mathbf{M}$ \cite{Nomura10}.
Our model Hamiltonian is hence:
\begin{equation}
\hat{H}=-i\hbar v_{F}\hat{\boldsymbol{\sigma}}\cdot\left(\boldsymbol{\nabla}\times\hat{\mathbf{z}}\right)+\Delta\hat{\boldsymbol{\sigma}}\cdot\mathbf{M},\label{H2d}
\end{equation}
where $v_{F}$ is the Fermi velocity of the Dirac fermions propagating
with momentum $\hbar\mathbf{k}$ measured relative to the $\Gamma$
point of the surface Brillouin zone. For $\mathrm{Bi}_{2}\mathrm{Te}_{3}$
the Fermi velocity is $v_{F}=4.3\times10^{5}\,\mathrm{m/s}$ \cite{Qi11}.
Here, $\hat{\boldsymbol{\sigma}}$ is the Pauli matrix operator and
$\Delta$ the proximity-induced exchange energy. Eq.~(\ref{H2d})
leads to the energy dispersion 
\begin{equation}
E_{ks}=s\sqrt{(\hbar v_{F}k_{x}+\Delta M_{y})^{2}+(\hbar v_{F}k_{y}-\Delta M_{x})^{2}+\left(\Delta M_{z}\right)^{2}},\label{E2d}
\end{equation}
where $s=\pm$ corresponds to the upper and lower bands. For an in-plane
exchange field we can rewrite Eq. (\ref{H2d}) as $\hat{H}=v_{F}\left(-i\hbar\boldsymbol{\nabla}-e\mathbf{A}\right)\cdot\left(\hat{\mathbf{z}}\times\hat{\boldsymbol{\sigma}}\right)+\Delta M_{z}\hat{\sigma}_{z}.$
The vector potential $\mathbf{A}=-\Delta/(ev_{F})\mathbf{M}\times\hat{\mathbf{z}}$
shifts the position of the Dirac point in the $(k_{x},k_{y})$- plane
and the electron charge is $-e$. A uniform and static $\mathbf{A}$
can be removed by the gauge transformation $(k_{x},k_{y})\rightarrow(q_{x}+eA_{x}/\hbar,q_{y}+eA_{y}/\hbar)$
and hence does not affect the physical observables. The energy dispersion
is then $E_{qs}=s\sqrt{(\hbar v_{F})^{2}(q_{x}^{2}+q_{y}^{2})+\left(\Delta M_{z}\right)^{2}}$
and eigenfunctions can be written as $\psi_{qs}=e^{i\mathbf{q}\cdot\mathbf{r}}|u_{qs}\rangle$
with 
\begin{equation}
|u_{q+}\rangle=\begin{pmatrix}\cos(\theta/2)\\
-ie^{i\phi}\sin(\theta/2)
\end{pmatrix},\ |u_{q-}\rangle=\begin{pmatrix}\sin(\theta/2)\\
ie^{i\phi}\cos(\theta/2)
\end{pmatrix},
\end{equation}
where $\cos\theta=\Delta M_{z}/|E_{qs}|$ and $\tan\phi=q_{y}/q_{x}$
determine the polar angle and the azimuth of the spinors on the Bloch
sphere.

The electron density of massless Dirac electrons relative to the neutrality
point reads 
\begin{equation}
n_{e}=\int_{0}^{E_{F}^{(0)}}dED_{0}(E)=\frac{\left(E_{F}^{(0)}\right)^{2}}{4\pi(\hbar v_{F})^{2}},\label{electron density}
\end{equation}
where $D_{0}(E)=\sum_{qs}\delta\left(E_{qs}-E\right)=E/[2\pi(\hbar v_{F})^{2}]$
is the density of states per unit area. $E_{F}^{(0)}=\hbar v_{F}\sqrt{4\pi n_{e}}$
is the Fermi energy for the gapless dispersion or in-plane magnetization.
When the electron density $n_{e}$ is kept constant under a rotating
magnetization, the Fermi energy of the gapped state reads $E_{F}(M_{z})=E_{F}^{(0)}\sqrt{1+\zeta^{2}M_{z}^{2}}$
with $\zeta=\Delta/E_{F}^{(0)}$.


\section{The Kubo formula}

\label{Kubo}

The MR is accessed in linear current response to an applied voltage
(Ohm's Law). Here we calculate electric dc conductivity in magnetized
2D Dirac electron system with nonmagnetic disorder by the Kubo formalism.
We assume that transport is limited by a randomly distributed disorder
potential 
\begin{equation}
\hat{V}(\mathbf{r})=V_{0}\sum_{i=1}^{N}\delta(\mathbf{r}-\mathbf{R}_{i})\label{normal impurity}
\end{equation}
that is weak and short-range Gaussian correlated $\langle\hat{V}(\mathbf{r}_{1})\hat{V}(\mathbf{r}_{2})\rangle_{\mathrm{imp}}=nV_{0}^{2}\delta(\mathbf{r}_{1}-\mathbf{r}_{2})$
with impurity concentration $n$ and (normal) scattering potential
$V_{0}$. In writing the impurity potentials as 2D
delta functions, we implicitly integrated over the envelope function
of the TI surface state thereby including bulk impurities close to
the interface. We focus on the dc conductivity of Zeeman-split 2D
Dirac electrons at zero temperature expressed in terms of the retarded
and advanced Green functions. This approach has previously been applied
to, e.g., the anomalous Hall effect (AHE) \cite{Dugaev05,Inoue06,Nunner07}
and the AMR \cite{Kato08} for spin-polarized 2D electrons with Rashba
SOC. In the diffusive transport regime, the Kubo formula for the dc
conductivity can be written as 
\begin{equation}
\sigma_{x\nu}=\frac{\hbar}{2\pi L^{2}}\operatorname{Tr}\left\langle \hat{\jmath}_{x}\hat{G}^{R}\hat{\jmath}_{\nu}\hat{G}^{A}\right\rangle _{\mathrm{imp}},\label{Kubo formula}
\end{equation}
where $L^{2}$ is the system area, $\hat{G}^{R(A)}(\epsilon)=\left(\epsilon\pm i0-\hat{H}-\hat{V}\right)^{-1}$
the retarded (advanced) Green function in the Pauli spin space. The
current operator reads $\hat{\mathbf{j}}=-e(-i/\hbar)[\hat{\mathbf{r}},\hat{H}]=-ev_{F}\hat{\mathbf{z}}\times\hat{\boldsymbol{\sigma}}$
. $\langle\cdots\rangle_{\mathrm{imp}}$ indicates an ensemble average
over random realizations of the impurity potential that we treat in
the Born approximation for the self-energy and the ladder approximation
for the current vertex \cite{Kato08}. The conductivity then reads
\begin{align}
\sigma_{x\nu} & \approx\frac{\hbar}{2\pi L^{2}}\operatorname{Tr}\left[\hat{\jmath}_{x}\langle\hat{G}^{R}\rangle\hat{\jmath}_{\nu}\langle\hat{G}^{A}\rangle\right]+\mathrm{Vertex\,correction}\nonumber \\
 & \equiv\frac{\hbar}{2\pi L^{2}}\operatorname{Tr}\left[\hat{\jmath}_{x}\langle\hat{G}^{R}\rangle\hat{J}_{\nu}\langle\hat{G}^{A}\rangle\right],\label{Kubo formula2}
\end{align}
where $\langle\hat{G}^{R(A)}\rangle$ is the averaged Green function
and $\hat{J}_{\nu}$ the corrected current vertex that includes the
diffuse scattering from impurities. The latter vanishes for short-range
impurity scattering in simple electron gases, but can be important
in the presence of impurity scattering, leading for instance to the
dephasing of the intrinsic spin Hall effect in the Rashba 2DEG \cite{Inoue04}

\subsection{Self-energy}

Here we calculate the averaged Green function by solving the Dyson
equation in the Born approximation as shown in Fig.~\ref{fig:Diagram}.
Hence, the averaged Green function can be written 
\begin{align}
\langle\hat{G}^{R(A)}\rangle & =\langle\left(z-\hat{H}-\hat{V}\right)^{-1}\rangle_{\mathrm{imp}}\nonumber \\
 & =\hat{G}_{0}^{R(A)}+\hat{G}_{0}^{R(A)}\hat{\Sigma}^{R(A)}\langle\hat{G}^{R(A)}\rangle
\end{align}
with $z=\epsilon\pm i0$. The solution to this equation is 
\begin{equation}
\langle\hat{G}^{R(A)}\rangle=\left((\hat{G}_{0}^{R(A)})^{-1}-\hat{\Sigma}^{R(A)}\right)^{-1}
\end{equation}
with the self-energy 
\begin{equation}
\hat{\Sigma}^{R(A)}=\langle\hat{V}\rangle_{\mathrm{imp}}+\langle\hat{V}\hat{G}_{0}^{R(A)}\hat{V}\rangle_{\mathrm{imp}}
\end{equation}
and the constant average $\langle\hat{V}\rangle_{\mathrm{imp}}$ is
absorbed in the Fermi energy $E_{F}$ in the following. In terms of
the unperturbed Green function 
\begin{equation}
\hat{G}_{0}^{R(A)}=\sum_{qs}\left(\epsilon-E_{qs}\pm i0\right)^{-1}|u_{qs}\rangle\langle u_{qs}|,
\end{equation}
\begin{align}
\langle\hat{V}\hat{G}_{0}^{R(A)}\hat{V}\rangle_{\mathrm{imp}} & =nV_{0}^{2}\int\frac{d^{2}\mathbf{q}}{(2\pi)^{2}}\hat{G}_{0}^{R(A)}\nonumber \\
 & =\mp i\frac{\hbar}{4\tau_{\mathrm{e}}}(1+\xi M_{z}\hat{\sigma}_{z}),\label{2nd self-energy}
\end{align}
where $1/\tau_{\mathrm{e}}=nV_{0}^{2}\int_{0}^{\infty}qdq\delta(E_{F}-E_{q+})/\hbar=2\pi nV_{0}^{2}D(E_{F})/\hbar=2\pi nV_{0}^{2}E_{F}/[h(\hbar v_{F})^{2}]$
denotes the elastic scattering rate and $\xi=\Delta/E_{F}=\zeta/\sqrt{1+\zeta^{2}M_{z}^{2}}$.
Eq.~(\ref{2nd self-energy}) shows how the self energy is modulated
by the magnetization direction. Hence, the averaged Green function
is 
\begin{equation}
\langle\hat{G}^{R(A)}\rangle=\frac{\varepsilon\pm i\Gamma_{0}+\hbar v_{F}\mathbf{q}\cdot(\hat{z}\times\hat{\boldsymbol{\sigma}})+(\Delta M_{z}\mp i\Gamma_{1})\hat{\sigma}_{z}}{(\varepsilon-E_{\mathbf{q}}^{+}\pm i\gamma^{+})(\varepsilon-E_{\mathbf{q}}^{-}\pm i\gamma^{-})},\label{averaged Green function}
\end{equation}
where $\Gamma_{0}=1/(4\tau_{\mathrm{e}})$, $\Gamma_{1}=\Gamma_{0}\cos\theta$,
and $\gamma^{\pm}=\Gamma_{0}(1\pm\cos^{2}\theta)$.

\subsection{Current vertex correction}

\begin{figure}[t]
\begin{centering}
\includegraphics[width=0.35\textwidth,angle=0]{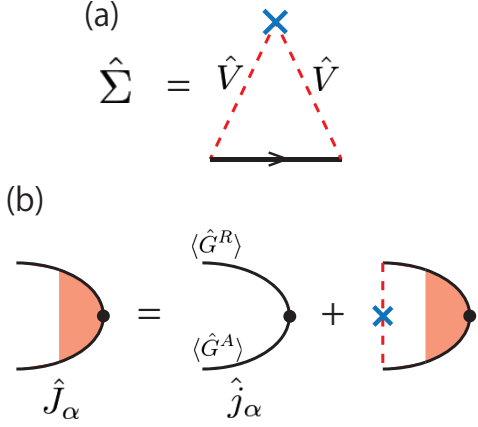} 
\par\end{centering}
\caption{(a) Self-energy diagram in the Born approximation. (b) The current
vertex correction is the geometric sum of ladder diagrams. }
\label{fig:Diagram}
\end{figure}

The vertex function in the Born approximation is represented by the
sum of all ladder diagrams in Fig.~\ref{fig:Diagram}. The self-consistent
Born approximation of the self-energy is consistent with the ladder
approximation to the vertex correction, while the first-order Born
approximation holds in the limit of weak disorder. This correspondence
has been confirmed for the AHE \cite{Sinitsyn06,Sinitsyn07}. Hence,
we treat the AMR within the ladder approximation and the first-order
Born approximation, which leads to an analytical formula for the conductivity
that agrees with the solution of the Boltzmann equation (see below).

The ladder-type vertex-corrected current operator $\hat{J}_{\nu}$
in Fig.~\ref{fig:Diagram} obeys the integral (Bethe-Salpeter) equation
\cite{Sinitsyn06,Sinitsyn07,Ado15,Ndiaye15} 
\begin{equation}
\hat{J}_{\nu}=\hat{\jmath}_{\nu}+nV_{0}^{2}\int\frac{d^{2}\mathbf{q}}{(2\pi)^{2}}\langle\hat{G}^{R}\rangle\hat{J}_{\nu}\langle\hat{G}^{A}\rangle.\label{VS eq}
\end{equation}
By iteration and Eq.~(\ref{averaged Green function}), the first-order
single-impurity vertex correction reads 
\begin{equation}
\hat{J}_{\nu}^{\left(1\right)}=nV_{0}^{2}\int\frac{d^{2}\mathbf{q}}{(2\pi)^{2}}\langle\hat{G}^{R}\rangle\hat{\jmath}_{\nu}\langle\hat{G}^{A}\rangle=-ev(\mp A\hat{\sigma}_{\nu}+B\hat{\sigma}_{\bar{\nu}})
\end{equation}
\,\ ($-$ for $\nu=x\,,\bar{\nu}=y$ and $+$ for $\nu=y\,,\bar{\nu}=x$)
with 
\begin{equation}
A=\frac{1-\xi^{2}M_{z}^{2}}{2(1+\xi^{2}M_{z}^{2})}\,,B=\frac{\hbar}{E_{F}\tau_{\mathrm{e}}}\frac{\xi M_{z}}{2(1+\xi^{2}M_{z}^{2})}.
\end{equation}
Expanding $\hat{J}_{\nu}=-ev_{F}\sum_{i}c_{\nu i}\hat{\sigma}_{i}$
in Eq.~(\ref{VS eq}) as 
\begin{equation}
\hat{J}_{\nu}=ev_{F}\left(\pm\hat{\sigma}_{\bar{\nu}}-nV_{0}^{2}\sum_{i}\int\frac{d^{2}\mathbf{q}}{(2\pi)^{2}}\langle\hat{G}^{R}\rangle c_{\nu i}\hat{\sigma}_{i}\langle\hat{G}^{A}\rangle\right),
\end{equation}
we find in the weak scattering limit 
\begin{equation}
\begin{pmatrix}c_{\nu x}\\
c_{\nu y}
\end{pmatrix}=\frac{1}{(1-A)^{2}}\begin{pmatrix}B & 1-A\\
-(1-A) & B
\end{pmatrix}\begin{pmatrix}\delta_{x\nu}\\
\delta_{y\nu}
\end{pmatrix}
\end{equation}
and $c_{\nu0}=c_{\nu z}=0$, where $\delta_{x\nu}$ and $\delta_{y\nu}$
are the Kronecker delta. For the limit of $\hbar/(E_{F}\tau_{\mathrm{e}})\ll1$,
the renormalized current vertex reads 
\begin{equation}
\begin{pmatrix}\hat{J}_{x}\\
\hat{J}_{y}
\end{pmatrix}=-ev_{F}\begin{pmatrix}-a & b\\
b & a
\end{pmatrix}\begin{pmatrix}\hat{\sigma}_{y}\\
\hat{\sigma}_{x}
\end{pmatrix}\label{renormalized current vertex}
\end{equation}
with 
\begin{align}
\begin{split}a & =c_{yx}=-c_{xy}=2\frac{1+\xi^{2}M_{z}^{2}}{1+3\xi^{2}M_{z}^{2}},\\
b & =c_{xx}=c_{yy}=2\frac{\hbar}{E_{F}\tau_{\mathrm{e}}}\frac{\xi M_{z}(1+\xi^{2}M_{z}^{2})}{(1+3\xi^{2}M_{z}^{2})^{2}}.\label{ab}
\end{split}
\end{align}
In the gapless limit of $\xi M_{z}\rightarrow0$ this reduces to $a=2$
and $b=0$.

\subsection{Longitudinal and transverse conductivities}

\begin{figure}[t]
\begin{centering}
\includegraphics[width=0.47\textwidth,angle=0]{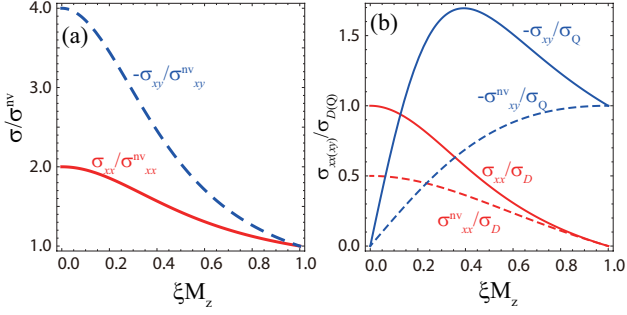} 
\par\end{centering}
\caption{(a) Ratio of the dc conductivities without ($\sigma_{x\alpha}^{\mathrm{nv}})$
and with vertex correction ($\sigma_{x\alpha})$ as a function of
$\xi M_{z}$. (b) The longitudinal and transverse conductivities without
and with vertex correction as a function of $\xi M_{z}$. $\sigma_{D}=\left(2e^{2}/h\right)\left(E_{F}\tau^{\mathrm{e}}/\hbar\right)$
is the longitudinal conductivity of two-dimensional massless Dirac
electrons without magnetic or exchange fields, while $\sigma_{\mathrm{Q}}=e^{2}/(2h)$.}
\label{fig:KuboVsnv}
\end{figure}

Inserting Eq.~(\ref{averaged Green function}) and Eq.~(\ref{renormalized current vertex})
into Eq.~(\ref{Kubo formula2}), 
\begin{align}
\begin{split}\sigma_{xx} & =a\sigma_{xx}^{\mathrm{nv}}+b\sigma_{xy}^{\mathrm{nv}},\\
\sigma_{xy} & =-b\sigma_{xx}^{\mathrm{nv}}+a\sigma_{xy}^{\mathrm{nv}}.\label{vc}
\end{split}
\end{align}
Here $\sigma_{x\nu}^{\mathrm{nv}}=\hbar\operatorname{Tr}\left[\hat{\jmath}_{x}\langle\hat{G}^{R}\rangle\hat{\jmath}_{\nu}\langle\hat{G}^{A}\rangle\right]/\left(2\pi L^{2}\right)$
are the longitudinal and transverse conductivities without vertex
correction (``bare bubbles\textquotedblright ):
\begin{align}
\begin{split}\sigma_{xx}^{\mathrm{nv}} & =\frac{e^{2}}{h}\frac{E_{F}\tau_{\mathrm{e}}}{\hbar}\frac{1-\xi^{2}M_{z}^{2}}{1+\xi^{2}M_{z}^{2}},\\
\sigma_{xy}^{\mathrm{nv}} & =-\frac{e^{2}}{2h}\frac{2\xi M_{z}}{1+\xi^{2}M_{z}^{2}}.\label{nv}
\end{split}
\end{align}
When the gap vanishes with $\xi M_{z}\rightarrow0$, the longitudinal
and transverse conductivities reduce to 
\begin{equation}
\sigma_{xx}^{\mathrm{nv}}=\frac{e^{2}}{h}\frac{E_{F}\tau_{\mathrm{e}}}{\hbar}
\end{equation}
and $\sigma_{xy}^{\mathrm{nv}}=0$. Below we show that $\sigma_{xx}^{\mathrm{nv}}$
is half of the full (vertex-corrected) result {[}Eq.~(\ref{gapless vc cond}){]}.
This discrepancy reflects the inherent anisotropy of the scattering
of Dirac fermions that affects the transport and relaxation times
even for short-range correlated scattering. Substituting Eqs.~(\ref{ab})
and (\ref{nv}) into Eq.~(\ref{vc}): 
\begin{align}
\sigma_{xx} & =2\frac{e^{2}}{h}\frac{E_{F}\tau_{\mathrm{e}}}{\hbar}\frac{1-\xi^{2}M_{z}^{2}}{1+3\xi^{2}M_{z}^{2}},\label{vcxx}\\
\sigma_{xy} & =-\frac{e^{2}}{2h}\xi M_{z}\frac{8(1+\xi^{2}M_{z}^{2})}{(1+3\xi^{2}M_{z}^{2})^{2}}.\label{vc cond}
\end{align}
For $\xi M_{z}\rightarrow0$ 
\begin{equation}
\sigma_{xx}=2\frac{e^{2}}{h}\frac{E_{F}\tau_{\mathrm{e}}}{\hbar}\equiv\frac{e^{2}}{h}\frac{E_{F}\tau}{2\hbar}\label{gapless vc cond}
\end{equation}
and $\sigma_{xy}=0$, where $\tau=4\tau_{\mathrm{e}}$ is the transport
relaxation time of massless Dirac electrons. $\sigma_{xx}(\xi M_{z}=0)$
is the longitudinal conductivity of nonmagnetic 2D massless Dirac
electrons \cite{Shon98,Culcer10}, which implies that the in-plane
exchange potential has no effect on electron transport as expected
from the gauge-field argument above. We here disregard the third order
``skew-scattering\textquotedblright \ term. Otherwise,
our $\sigma_{xy}\left(M_{z}\right)$ agrees with previous results
\cite{Sinitsyn06,Ado15,Culcer11,Sakai14}. $\sigma_{xx}\left(M_{z}\right)$
has been derived in \cite{Sakai14}. Fig.~\ref{fig:KuboVsnv}\,(a)
shows the ratio of the dc conductivities without and with the ladder-vertex
correction as a function of $\xi M_{z}$, while Fig.~\ref{fig:KuboVsnv}\,(b)
is a plot of the $\xi M_{z}$-dependence of the conductivities $\sigma_{x\nu}$
and $\sigma_{x\nu}^{\mathrm{nv}}$. When the electron density (Eq.~(\ref{electron density}))
is kept constant for all $\mathbf{M}$, the longitudinal conductivity
becomes 
\begin{equation}
\sigma_{xx}=2\frac{e^{2}}{h}\frac{E_{F}^{(0)}\tau_{\mathrm{e}}^{(0)}}{\hbar}\frac{1}{1+4\zeta^{2}M_{z}^{2}},
\end{equation}
where $1/\left(E_{F}^{(0)}\tau_{\mathrm{e}}^{(0)}\right)=2\pi nV_{0}^{2}/[h(\hbar v_{F})^{2}]=1/\left(E_{F}\tau_{\mathrm{e}}\right)$.
Hence, to leading order in $\left(\zeta M_{z}\right)^{2}$ the MRs
for constant density or Fermi energy are the same.

\subsection{Parameter dependence}

\begin{figure*}[t]
\begin{centering}
\includegraphics[width=0.95\textwidth,angle=0]{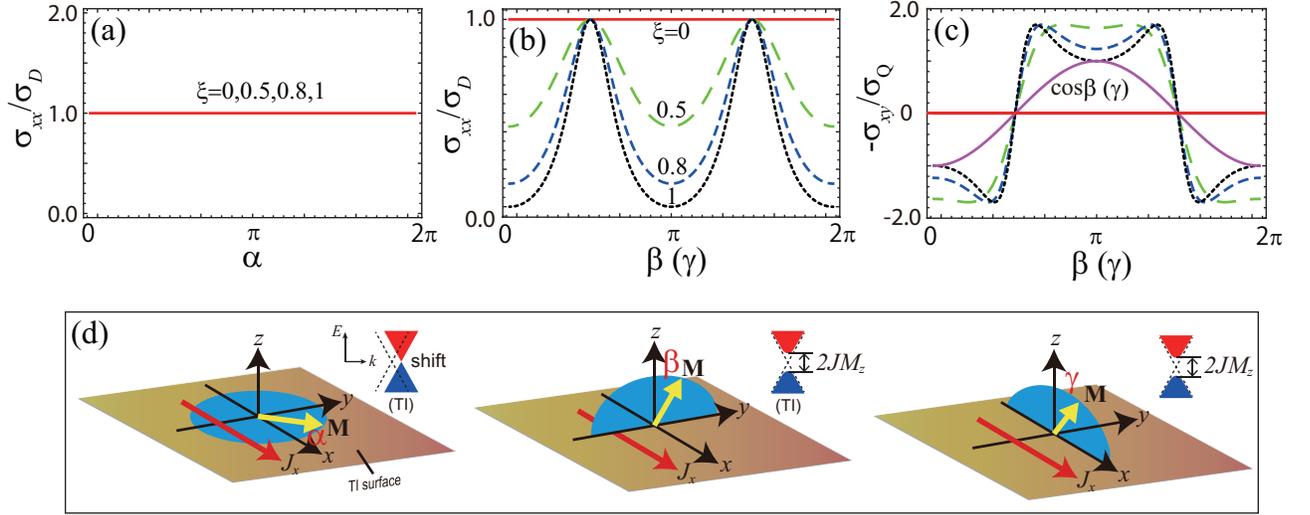} 
\par\end{centering}
\caption{Calculated conductivities in the TI$|$FI bilayer as a function of
magnetization angles $\alpha$, $\beta$, and $\gamma$ (shown in
(d)) and for different ratios $\xi=J/E_{F}$. (a), (b) show the longitudinal
conductivity and (c) the transverse (Hall) conductivity. Each subplot
in (d) shows a different configuration and associated band structure
of the surface state. $\sigma_{D}$\textit{ }and $\sigma_{Q}\ $are
defined in Fig.\ref{fig:KuboVsnv}. The dependence on the angles $\beta$
and $\gamma$ is here the same. The $\cos\beta\left(\gamma\right)$
function is plotted in (c) for reference.}
\label{fig:MRcurve}
\end{figure*}

Figure~\ref{fig:MRcurve}\,(a) and (b) show the longitudinal conductivity
$\sigma_{xx}$ as a function of angle $\alpha$, $\beta$, and $\gamma$
of the FI magnetization in $x$-$y$, $y$-$z$, and $x$-$z$ planes,
respectively., while Fig.~\ref{fig:MRcurve}\,(c) shows the transverse
conductivity $\sigma_{xy}$ for different Fermi energies $E_{F}$.
The calculated results for $\sigma_{xx}=\sigma_{xx}(M_{z}^{2})$ are
very similar to those computed for magnetically doped TIs \cite{Sabzalipour15}.
The inset in each magnetization rotation in Fig.~\ref{fig:MRcurve}\,(d)
illustrate the band structure: When the magnetization is in-plane,
the bands are rigidly shifted in the $k_{x},k_{y}$-plane, which does
not affect the MR. In contrast, an out-of-plane magnetization opens
a gap that suppresses the longitudinal conductivity.

\section{Boltzmann transport theory}

\label{Boltzmann}

\subsection{Transport time}

Here we employ the Boltzmann equation to calculate the electric dc
conductivity of magnetized 2D massless Dirac electron system with
(initially) nonmagnetic disorder and arrive at results that are identical
with those from the Kubo formalism in the previous section and
Ref.\onlinecite{Sakai14}. We show that the
in-scattering term of the collision integral in Boltzmann theory is
significant and equivalent with the current-vertex correction in linear
response theory (see Sec.~\ref{Kubo}). Sufficiently far from the
Dirac point the impurity scattering can be treated by the Born approximation
\cite{Adam09}. The non-equilibrium distribution function $f(\mathbf{q})$
in the presence of a uniform external electric field $\mathbf{E}$
is governed by the linearized Boltzmann equation 
\begin{equation}
-e\left(-\frac{\partial f^{\left(0\right)}}{\partial E_{qs}}\right)\mathbf{v}_{qs}\cdot\mathbf{E}=\left(\frac{\partial f}{\partial t}\right)_{\mathrm{scat}},\label{Boltz eq}
\end{equation}
where $\mathbf{v}_{qs}=\nabla_{\mathbf{q}}E_{qs}/\hbar$ is the group
velocity and $f^{(0)}(\mathbf{q})$ the equilibrium Fermi-Dirac distribution
function. The collision term on the right hand side is affected by
in- and out-scattering of the state with wave vector $\mathbf{q}$
\begin{equation}
\left(\frac{\partial f}{\partial t}\right)_{\mathrm{scat}}=\frac{1}{L^{2}}\sum_{\mathbf{q}^{\prime}}W_{\mathbf{q,q^{\prime}}}\left(f(\mathbf{q}^{\prime})-f(\mathbf{q})\right),\label{Scat term}
\end{equation}
where $W_{\mathbf{q,q^{\prime}}}$ is the transition probability between
$\mathbf{q}$ and $\mathbf{q}^{\prime}$ states. Elastic impurity
scattering implies $|\mathbf{q}|=|\mathbf{q}^{\prime}|$. By Fermi's
golden rule: $W_{\mathbf{q,q^{\prime}}}=(2\pi/\hbar)|T_{\mathbf{q,q^{\prime}}}|^{2}\delta\left(E_{qs}-E_{q^{\prime}s}\right)$
with $T$-matrix element $T_{\mathbf{q,q^{\prime}}}$ for scattering
from $\mathbf{q}$ to $\mathbf{q}^{\prime}$. The transition rate
can be expressed in terms of the disorder potential Eq.~(\ref{normal impurity}).
Combining Eqs.~(\ref{Boltz eq}) and (\ref{Scat term}), the transport
time of Dirac electrons in the Born approximation reads 
\begin{equation}
\frac{1}{\tau(\mathbf{q})}=\int\frac{d^{2}\mathbf{q}^{\prime}}{(2\pi)^{2}}W_{\mathbf{q,q^{\prime}}}\left(1-\cos\left(\mathbf{q^{\prime},q}\right)\right),
\end{equation}
where the in-scattering term contributes to the factor $\cos\left(\mathbf{q^{\prime},q}\right)=\mathbf{q}^{\prime}\cdot\mathbf{q}/q^{2}=\cos(\phi-\phi^{\prime})$
that is associated with the ladder-vertex correction in the Kubo theory
\cite{Rickayzen,Sinitsyn07}.

To lowest order in the scattering potential (thereby disregarding
skew scattering as above) the transition probability in the upper
band reads 
\begin{align}
|T_{\mathbf{q,q^{\prime}}}|^{2} & \approx\langle|\langle u_{q^{\prime}+}|\hat{V}|u_{q+}\rangle|^{2}\rangle_{\mathrm{imp}}\nonumber \\
 & =nV_{0}^{2}|\langle u_{q^{\prime}+}|u_{q+}\rangle|^{2}\nonumber \\
 & =nV_{0}^{2}\left(1-\sin^{2}\theta\sin^{2}\frac{\phi-\phi^{\prime}}{2}\right),\label{Tqq}
\end{align}
leading to the electron transport relaxation time 
\begin{align}
\frac{1}{\tau} & =\int\frac{d^{2}\mathbf{q}^{\prime}}{(2\pi)^{2}}\frac{2\pi}{\hbar}|T_{\mathbf{q,q^{\prime}}}|^{2}\left(1-\cos(\phi-\phi^{\prime})\right)\delta\left(E_{F}-E_{q^{\prime}+}\right)\nonumber \\
 & =\frac{1}{4\tau_{\mathrm{e}}}\left(1+3\xi^{2}M_{z}^{2}\right).\label{transport time}
\end{align}
This result reduces to the transport relaxation time of massless Dirac
electrons $\tau=4\tau_{\mathrm{e}}$ for $\xi M_{z}\rightarrow0$.
From Eq.~(\ref{2nd self-energy}), the transport time without the
vertex correction is 
\begin{align}
\frac{1}{\tau^{\mathrm{nv}}} & =-2\operatorname{Im}\hat{\Sigma}^{R}\nonumber \\
 & =\int\frac{d^{2}\mathbf{q}^{\prime}}{(2\pi)^{2}}\frac{2\pi}{\hbar}|T_{\mathbf{q,q^{\prime}}}|^{2}\delta\left(E_{F}-E_{q^{\prime}+}\right)\nonumber \\
 & =\frac{1}{4\tau_{\mathrm{e}}}2\left(1+\xi^{2}M_{z}^{2}\right),
\end{align}
while the transport time with in scattering is expressed as Eq.~(\ref{transport time}).
On the other hand, Eq.~(\ref{VS eq}) gives a corrected velocity
(or current) of the form $v_{x}=sav\sin\theta\cos\phi$ with $a=\tau/\tau^{\mathrm{nv}}$,
which directly relates the ladder-vertex correction in the Kubo theory
with the Boltzmann transport time \cite{Sinitsyn07}. Therefore, we
can confirm that the ladder-vertex correction and in scattering terms
both renormalizes the velocity in the same way.

\subsection{Longitudinal and transverse conductivities}

Here we calculate the charge current $\mathbf{J}_{c}$ driven by an
in-plane electric field as a function of the exchange field direction
$\mathbf{M}$ as shown in Fig.~\ref{fig:system2d}. The corresponding
nonequilibrium distribution function is $f(\mathbf{q})=f^{(0)}(\mathbf{q})+g(\mathbf{q})=f^{(0)}(\mathbf{q})+e\tau\left(\partial f^{(0)}/\partial E_{qs}\right)\mathbf{v}_{qs}\cdot\mathbf{E}$,
where at zero temperature $f^{\left(0\right)}(\mathbf{q})\approx\theta(E_{F}-E_{qs})$.
To leading order in $\mathbf{E}=E_{x}\hat{\mathbf{x}}$ 
\begin{equation}
\sigma_{ix}=\frac{\mathbf{J}_{c}\cdot\hat{\mathbf{e}}_{i}}{E_{x}}=\frac{-e}{E_{x}}\sum_{s}\int\frac{d^{2}\mathbf{q}}{(2\pi)^{2}}g(\mathbf{q})\mathbf{v}_{qs}\cdot\hat{\mathbf{e}}_{i},\label{Jxx}
\end{equation}
where $\hat{\mathbf{e}}_{i}=\hat{\mathbf{x}},\,\hat{\mathbf{y}}$
and the electron velocity $\mathbf{v}_{qs}=\langle\psi_{qs}|\hat{\mathbf{v}}|\psi_{qs}\rangle$
is the expectation value of the velocity operator $\hat{\mathbf{v}}=(-i/\hbar)[\hat{\mathbf{r}},\hat{H}]=v_{F}\hat{\mathbf{z}}\times\hat{\boldsymbol{\sigma}}$
or group velocity $\mathbf{v}_{qs}=v_{x}\hat{\mathbf{x}}+v_{y}\hat{\mathbf{y}}$
with $v_{x}=sv_{F}\sin\theta\cos\phi$ and $v_{y}=sv_{F}\sin\theta\sin\phi$.
When the Fermi energy is above the gap, i.e., $E_{F}>\Delta$, the
longitudinal and transverse conductivities are 
\begin{align}
\sigma_{xx} & =2\frac{e^{2}}{h}\frac{E_{F}\tau_{\mathrm{e}}}{\hbar}\frac{1-\xi^{2}M_{z}^{2}}{1+3\xi^{2}M_{z}^{2}}.\label{sxxl}\\
\sigma_{xy} & =0.\label{sxyl}
\end{align}
In contrast to the linear response result Eq. (\ref{vc cond}), $\sigma_{xy}$
vanishes, because intrinsic (Berry phase) and side-jump scattering
contributions are not included in Eq. (\ref{Boltz eq}). Sinitsyn
\textit{et al}. \cite{Sinitsyn07} demonstrated that and how the Boltzmann
equation can be repaired to recover the diagrammatic results for the
AHE. We can disregard this complication for the MR, the focus of the
present study, since the intrinsic mechanism and the side jump scattering
(to leading order) do not contribute to longitudinal transport.

\subsection{Current-induced spin polarization}

Here we discuss the conductivities derived above
in terms of current-induced torque to the magnetization \cite{Sakai14}
in the metallic regime ($E_{F}>\Delta$) \cite{Ndiaye15,Garate10}.
The electric-field-driven non-equilibrium spin density or Edelstein
effect \cite{Edelstein90,Inoue03} can be expressed by the Kubo formula
as well as by Boltzmann theory. For 2D massless Dirac electrons, the
charge current is proportional to the spin operator, as can be seen
from $\hat{\mathbf{j}}=-e\hat{\mathbf{v}}=-ev_{F}\hat{\mathbf{z}}\times\hat{\boldsymbol{\sigma}}$.
Therefore, a nonzero steady-state charge current implies a finite
spin density that can easily be found by multiplying the charge current
by $-\hbar/(2ev_{F})$, yielding 
\begin{equation}
\langle\mathbf{s}\rangle=\sum_{s}\int\frac{d^{2}\mathbf{q}}{(2\pi)^{2}}g(\mathbf{q})\mathbf{s}(q,s),\label{sxy}
\end{equation}
where $\mathbf{s}(q,s)=(\hbar/2)\langle\psi_{qs}|\hat{\boldsymbol{\sigma}}_{\perp}|\psi_{qs}\rangle=-\hbar/(2ev_{F})\hat{\mathbf{z}}\times\mathbf{v}(q,s)$
with $\hat{\boldsymbol{\sigma}}_{\perp}=(\hat{\sigma}_{x},\hat{\sigma}_{y})$,
i.e., for $\mathbf{E}=E_{x}\hat{\mathbf{x}}$ 
\begin{equation}
\langle s_{x}\rangle=0,\,\langle s_{y}\rangle=-\frac{\hbar}{2}\frac{2eE_{x}}{hv_{F}}\frac{E_{F}\tau_{\mathrm{e}}}{\hbar}\frac{1-\xi^{2}M_{z}^{2}}{1+3\xi^{2}M_{z}^{2}}.\label{Sxy}
\end{equation}
The current-induced spin polarization is therefore not affected by
an in-plane magnetization in spite of the exchange interaction in
Eq.~(\ref{H2d}). Finite functional derivatives of the total energy
as a function of $\mathbf{M}$ are equivalent to effective fields
acting on the magnetization. From the exchange energy $E_{\mathrm{ex}}=\Delta\langle\mathbf{s}\rangle\cdot\mathbf{M,}$
we can compute the (field-like) $\mathbf{T}=-\gamma\left(\delta E_{\mathrm{ex}}/\delta\mathbf{M}\right)\times\mathbf{M}$
$=-\gamma\Delta\langle s_{y}\rangle\hat{\mathbf{y}}\times\mathbf{M}$,
where $\gamma$ is the gyromagnetic ratio. When $\mathbf{M}\Vert\mathbf{E}$,
the torque strives to rotate the magnetization \textit{out-of-plane,}
while it vanishes when $\mathbf{M}\Vert\hat{\mathbf{y}}$, just like
the torques induced by the spin Hall effect in metallic conductors.
However, the electric resistance is not affected because there is
no \textit{in-plane} (antidamping-like) torque. The $M_{z}$ dependence
of Eqs. (\ref{Sxy}) and (\ref{sxxl}) is the same, which is another
consequence of the spin-momentum locking in the Dirac electron system.

\section{Magnetic impurities}

\label{magnetic impurities}

We assumed above nonmagnetic scattering which might not be a good
representation of the TI
$\vert$
FI interface. Any roughness of this interface is likely to introduce
magnetic disorder on the TI surface that can be modeled by randomly
distributed magnetic impurities of spin $S$ with direction given
by the unit vector $\mathbf{S}_{\mathrm{m},i}$ with index $i$ at
positions $\mathbf{R}_{i}$ and scattering potential 
\begin{equation}
\hat{V}_{\mathrm{m}}(\mathbf{r})=V_{\mathrm{m}}\sum_{i=1}^{N_{{\rm m}}}\hat{\boldsymbol{\sigma}}\cdot\mathbf{S}_{\mathrm{m},i}\delta(\mathbf{r}-\mathbf{R}_{i}),\label{Vm}
\end{equation}
where $V_{\mathrm{m}}=J_{\mathrm{m}}S$ is the interaction strength
between the conduction electrons and the local moments with a magnitude
$S$ and exchange constant $J_{\mathrm{m}}.$ Since these impurities
are coupled to the FI magnetization, a large fraction is likely to
be parallel to $\mathbf{M.}$ TI surface states with magnetic impurities
(but without proximity ferromagnets) display various phases as a function
of the impurity concentration and temperature (here $T=0$) \cite{Liu09,Ochoa15}.
For example, a phase transition from a paramagnetic to an out-of-plane
ferromagnetic phase can be induced by increasing the impurity concentration.
Only when the Ruderman-Kittel-Kasuya-Yosida (RKKY) interaction among impurity spins is overcome by
the exchange interaction from the attached FI, the ferromagnetic phase
becomes aligned to the FI magnetization.

Next we calculate the conductivity in the presence
of magnetic impurities modeled by Eq.~(\ref{Vm}). Since we found
in previous sections that Kubo and Boltzmann theories give identical
results for the conductivities, we use the latter (and simpler) method
in the following.

\subsection{Magnetic impurity aligned to $\mathbf{M}$}

First, the magnetic impurities are assumed to be aligned such that
$\mathbf{S}_{\mathrm{m}}^{i}=\mathbf{M}$. Hence, the scattering potential
can be simplified to $\hat{V}_{\mathrm{m}}(\mathbf{r})=V_{\mathrm{m}}\hat{\boldsymbol{\sigma}}\cdot\mathbf{M}\sum_{i}\delta(\mathbf{r}-\mathbf{R}_{i}).$
The impurity moments contribute an exchange potential $\left\langle \hat{V}_{\mathrm{m}}\right\rangle =\Delta_{\mathrm{m}}M_{z}\hat{\sigma}_{z}$
to the the surface electrons, where $\Delta_{\mathrm{m}}=n_{\mathrm{m}}V_{\mathrm{m}}$
\cite{Nomura11} which can be added to the proximity exchange as $\tilde{\Delta}=\Delta+\Delta_{\mathrm{m}}$.

The transition probabilities in the upper band are then 
\begin{widetext}
\begin{align}
|T_{\mathbf{q,q^{\prime}}}^{(\mathrm{m})}|^{2}\approx \langle|\langle u_{q^{\prime}+}|\hat{V}_{\mathrm{m}}|u_{q+}\rangle|^{2}\rangle_{\mathrm{imp}}
= & n_{\mathrm{m}}V_{\mathrm{m}}^{2}|\langle u_{q^{\prime}+}|(\hat{\sigma}_{x}M_{x}+\hat{\sigma}_{y}M_{y}+\hat{\sigma}_{z}M_{z})|u_{q+}\rangle|^{2}\nonumber \\
= & n_{\mathrm{m}}V_{\mathrm{m}}^{2}\sin^{2}\theta\left(M_{x}^{2}\sin^{2}\frac{\phi+\phi^{\prime}}{2}+M_{y}^{2}\cos^{2}\frac{\phi+\phi^{\prime}}{2}-M_{x}M_{y}\sin(\phi+\phi^{\prime})\right)\nonumber \\
 & +n_{\mathrm{m}}V_{\mathrm{m}}^{2}M_{z}\cos\theta\sin\theta\left[M_{x}\left(\sin\phi+\sin\phi^{\prime}\right)-M_{y}\left(\cos\phi+\cos\phi^{\prime}\right)\right]\nonumber \\
 & +n_{\mathrm{m}}V_{\mathrm{m}}^{2}M_{z}^{2}\left(1-\sin^{2}\theta\cos^{2}\frac{\phi-\phi^{\prime}}{2}\right).
\end{align}
\end{widetext}

The associated electron transport time becomes 
\begin{align}
\frac{1}{\tau_{\mathrm{m}}(\mathbf{q})} & =\int\frac{d^{2}\mathbf{q}^{\prime}}{(2\pi)^{2}}\frac{2\pi}{\hbar}|T_{\mathbf{q,q^{\prime}}}^{(\mathrm{m})}|^{2}\left(1-\cos(\phi-\phi^{\prime})\right)\delta\left(E_{F}-E_{q^{\prime}+}\right)\nonumber \\
 & =\frac{1}{4\tau_{\mathrm{m}}^{\mathrm{e}}}\left(1-M_{z}^{2}\right)\left(1-\tilde{\xi}^{2}M_{z}^{2}\right)\left(2+\cos2(\phi_{M}-\phi)\right)\nonumber \\
 & +\frac{1}{4\tau_{\mathrm{m}}^{\mathrm{e}}}2\tilde{\xi}M_{z}^{2}\sqrt{(1-M_{z}^{2})(1-\tilde{\xi}^{2}M_{z}^{2})}\sin(\phi_{M}-\phi)\nonumber \\
 & +\frac{1}{4\tau_{\mathrm{m}}^{\mathrm{e}}}M_{z}^{2}\left(3+\tilde{\xi}^{2}M_{z}^{2}\right),\label{taum}
\end{align}
where 
\begin{align}
\frac{\hbar}{\tau_{\mathrm{m}}^{\mathrm{e}}}=n_{\mathrm{m}}V_{\mathrm{m}}^{2}\int_{0}^{\infty}qdq\delta(E_{F}-E_{q+})=2\pi n_{\mathrm{m}}V_{\mathrm{m}}^{2}D(E_{F})
\end{align}
defines the elastic scattering rate by magnetic impurities, $\tilde{\xi}=\tilde{\Delta}/E_{F}$,
and $\phi_{\mathrm{M}}\left(=\alpha\right)$ is the polar angle of
the magnetization. The $\mathbf{q}$-dependence is caused by spin-flip
scattering due to the in-plane magnetic impurities that contribute
to the transport relaxation through the spin-momentum locking.

\begin{figure}[t]
\begin{centering}
\includegraphics[width=0.4\textwidth,angle=0]{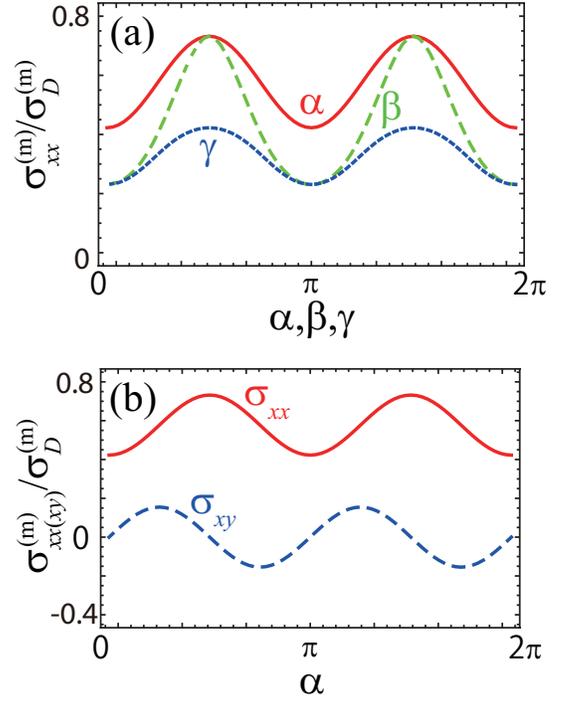} 
\par\end{centering}
\caption{(a) MR curves for a TI$|$FI bilayer as a function of magnetization
angles $\alpha$, $\beta$, and $\gamma$ (defined in Fig.~\ref{fig:MRcurve}
(d)) for $\tilde{\xi}=0.5$. $\sigma_{D}^{(\mathrm{m})}=\left(2e^{2}/h\right)\left(E_{F}\tau_{\mathrm{m}}^{\mathrm{e}}/\hbar\right)$
is the longitudinal conductivity limited by magnetic impurities. (b)
Longitudinal and transverse conductivities as a function of in-plane
magnetization angles $\alpha$.}
\label{fig:MRcurve2}
\end{figure}

To leading order in $\mathbf{E}=E_{x}\mathbf{\hat{x}}$ 
\begin{align}
\sigma_{ij}^{\left(\mathrm{m}\right)} & =e^{2}\sum_{s}\int\frac{d^{2}\mathbf{q}}{(2\pi)^{2}}\delta\left(E_{F}-E_{qs}\right)\tau_{\mathrm{m}}(\mathbf{q})v_{i}v_{j}\nonumber \\
 & =2\frac{e^{2}}{h}\frac{E_{F}\tau_{\mathrm{m}}^{\mathrm{e}}}{\hbar}\int_{0}^{2\pi}\frac{d\phi}{2\pi}\frac{F_{ij}(\phi)}{A-B\sin(\phi_{\mathrm{M}}-\phi)+\cos2(\phi_{\mathrm{M}}-\phi)},\label{Jxxm}
\end{align}
where $F_{xx}(\phi)=1+\cos2\phi$, $F_{xy}(\phi)=\sin2\phi$, $A=2+M_{z}^{2}(3+\tilde{\xi}^{2}M_{z}^{2})/[(1-M_{z}^{2})(1-\tilde{\xi}^{2}M_{z}^{2})]$,
and $B=2\tilde{\xi}M_{z}^{2}/\sqrt{(1-M_{z}^{2})(1-\tilde{\xi}^{2}M_{z}^{2})}$.
In the absence of out-of-plane spin components $M_{z}=0\,(A=2,B=0)$:
\begin{align}
\sigma_{xx}^{\left(\mathrm{m}\right)} & =2\frac{e^{2}}{h}\frac{E_{F}\tau_{\mathrm{m}}^{\mathrm{e}}}{\hbar}\frac{\sqrt{3}}{3}\left[1-\left(2-\sqrt{3}\right)\left(M_{x}^{2}-M_{y}^{2}\right)\right],\label{Jxxmxx}\\
\sigma_{xy}^{\left(\mathrm{m}\right)} & =2\frac{e^{2}}{h}\frac{E_{F}\tau_{\mathrm{m}}^{\mathrm{e}}}{\hbar}\frac{2\sqrt{3}}{3}\left(2-\sqrt{3}\right)M_{x}M_{y}.\label{Jxxmxy}
\end{align}
In contrast to the homogeneous proximity effect, a disordered in-plane
exchange potential generates an MR with periodicity $\pi$ as well
as a planar Hall effect. This result agrees with the AMR computed
for a single magnetic impurity by first principles \cite{Narayan15}
and explains the experimental in-plane MR of TI/FI bilayers \cite{Banerjee15}
as well as magnetically doped TIs \cite{Kandala15}. The reported
in-plane unidirectional MR (with periodicity $2\pi$) in the TI/magnetically-doped-TI
bilayer \cite{Yasuda16} is nonlinear (proportional to the applied
current), therefore beyond the linear response treatment here\textit{.}
In Fig.~\ref{fig:MRcurve2} (b), the planar Hall conductivity is
plotted as a function of in-plane magnetization angle $\alpha$ that
displays both AMR and SMR character. The planar Hall angle 
\begin{equation}
\theta_{\mathrm{PHE}}\equiv\frac{\sigma_{xy}^{\left(\mathrm{m}\right)}}{\sigma_{xx}^{\left(\mathrm{m}\right)}}=\frac{\sin2\alpha}{\left(2+\sqrt{3}\right)-\cos2\alpha}\leq0.28
\end{equation}
is maximal for $\alpha=\tan^{-1}(3^{-1/4})=37.2^{\circ}$. This Hall
angle for diffuse transport is much smaller than that predicted in
a ballistic transport model \cite{Scharf16}.

In Fig.~\ref{fig:MRcurve2} we plot the conductivity Eq.~(\ref{Jxxm})
as a function of the angles $\alpha,\beta,\gamma$ as defined in Fig.
\ref{fig:MRcurve}, where $\alpha=\phi_{\mathrm{M}}$ is the in-plane
angle, while $\beta,\gamma$ are out-of-plane angles $\theta_{\mathrm{M}}$
for $\alpha=\pi/2$ and $\alpha=0$, respectively.\textit{ }A sizable
MR with two-fold symmetry in all three orthogonal planes is shown
in Fig.~\ref{fig:MRcurve} (d). In Fig.~\ref{fig:MRcurve2} (a),
the conductivity $\sigma_{xx}$ {[}see Eq.~(\ref{Jxxmxy}){]} depends
only on the in-plane magnetization angle $\phi_{\mathrm{M}}$, which
can be explained in terms of electron scattering by the magnetic impurities
under spin-momentum locking of the Dirac electrons. When the magnetic
impurities are aligned with $\mathbf{M}$, electron scattering is
affected by the magnetization of the FI and the transport time is
modified. In contrast to the MR by nonmagnetic impurities, the out-of-plane
conductivities $\sigma_{xx}(\beta)$ and $\sigma_{xx}(\gamma)$ differ,
which again reflects the spin-momentum locking of Dirac electrons,
i.e., fixing the spin direction to $0\,(\pi/2)$ leads to different
momentum relaxations.

When the magnetic impurity concentration becomes higher, the impurity
spin direction may be locked perpendicular to the plane by the RKKY
coupling mediated by the Dirac electrons, which is stabilized by the
induced gap \cite{Liu09,Ochoa15}. Weak applied magnetic fields then
control only the direction of FI magnetization; only sufficiently
strong magnetic field may also rotate the orientation of the impurity
magnetization from the $z$-direction. In the former regime $\hat{V}_{\mathrm{m}}(\mathbf{r})=V_{\mathrm{m}}\hat{\sigma}_{z}\sum_{i}\delta(\mathbf{r}-\mathbf{R}_{i})$
and $\left\langle \hat{V}_{\mathrm{m}}\right\rangle =\Delta_{\mathrm{m}}\hat{\sigma}_{z}$
with upper band transition probabilities 
\begin{equation}
\left\vert T_{\mathbf{q},\mathbf{q}^{\prime}}^{(\mathrm{z})}\right\vert ^{2}=n_{\mathrm{m}}V_{\mathrm{m}}^{2}\left(1-\sin^{2}\theta\cos^{2}\frac{\phi-\phi^{\prime}}{2}\right),
\end{equation}
where that lead to thee following transport time and longitudinal
conductivity
\begin{equation}
\frac{1}{\tau_{z}}=\frac{1}{4\tau_{\mathrm{m}}^{e}}\left(3+\xi_{\mathrm{m}}^{2}+\xi^{2}M_{z}^{2}\right)
\end{equation}
\begin{equation}
\sigma_{xx}^{\left(\mathrm{z}\right)}=2\frac{e^{2}}{h}\frac{E_{F}\tau_{\mathrm{m}}^{\mathrm{e}}}{\hbar}\frac{1-\xi_{\mathrm{m}}^{2}-\xi^{2}M_{z}^{2}}{3+\xi_{\mathrm{m}}^{2}+\xi^{2}M_{z}^{2}}.
\end{equation}
with $\xi_{\mathrm{m}}=\Delta_{\mathrm{m}}/E_{F}.$ The conductivity
is now reduced because by scattering at a magnetic impurity the electron
acquires a phase shift $e^{i(\phi+\pi)}$ that enhances back scattering.

\subsection{Paramagnetic impurities}

When the magnetic impurities are paramagnetic, $\mathbf{S}_{\mathrm{m}}^{i}=\mathbf{S}_{\mathrm{m}}$
and assuming that the magnetic fields do not significantly polarize
the moments $\langle\mathbf{S}_{\mathrm{m}}\rangle=0$, the scattering
potential is reduced to $\hat{V}_{\mathrm{m}}(\mathbf{r})=V_{\mathrm{m}}\sum_{i}\hat{\boldsymbol{\sigma}}\cdot\mathbf{S}_{\mathrm{m},i}\delta(\mathbf{r}-\mathbf{R}_{i}).$
The transition amplitude in the upper band becomes 
\begin{align}
|T_{\mathbf{q,q^{\prime}}}^{(\mathrm{p})}|^{2} & \approx\langle|\langle u_{q^{\prime}+}|\hat{V}_{\mathrm{m}}|u_{q+}\rangle|^{2}\rangle_{\mathrm{imp}}\nonumber \\
 & =n_{\mathrm{m}}V_{\mathrm{m}}^{2}|\langle u_{q^{\prime}+}|\hat{\boldsymbol{\sigma}}\cdot\mathbf{S}_{\mathrm{m}}|u_{q+}\rangle|^{2}\nonumber \\
 & =n_{\mathrm{m}}V_{\mathrm{m}}^{2}\frac{1}{3}S^{2}\left(1+\sin^{2}\theta\sin^{2}\frac{\phi-\phi^{\prime}}{2}\right).
\end{align}
The electron transport time is 
\begin{equation}
\frac{1}{\tau_{\mathrm{p}}}=\frac{1}{12\tau_{\mathrm{m}}^{\mathrm{e}}}S^{2}\left(5-3\xi^{2}M_{z}^{2}\right).
\end{equation}
An in-plane component $(S_{\mathrm{m},i})_{x(y)}$ contributes to
the transport relaxation. To leading order in $\mathbf{E}=E_{x}\hat{\mathbf{x}}$,
\begin{equation}
\sigma_{xx}^{\left(\mathrm{p}\right)}=2\frac{e^{2}}{h}\frac{E_{F}\tau_{\mathrm{m}}^{\mathrm{e}}}{\hbar S^{2}}\frac{3(1-\xi^{2}M_{z}^{2})}{5-3\xi^{2}M_{z}^{2}}.
\end{equation}
The $z$-component of the magnetic impurity contributes a scattering
phase shift $e^{i(\phi+\pi)}$ and thereby additional back scattering.
Moreover, the $x,y$-components of the magnetic impurity locally break
the time-reversal symmetry on the TI-surface and allow back scattering,
which is weaker than for the polarized impurities, however.

\subsection{In-plane magnetoresistance }

\begin{figure}[t]
\begin{centering}
\includegraphics[width=0.45\textwidth,angle=0]{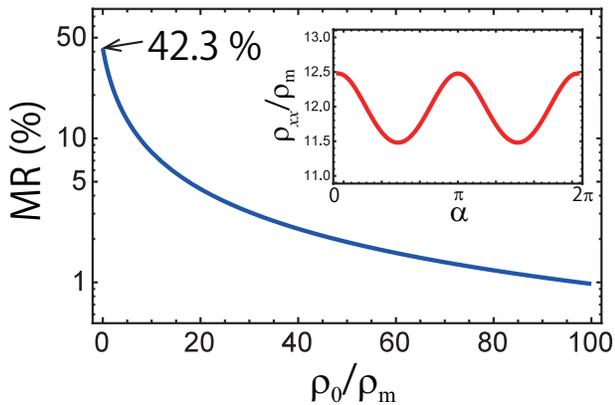} 
\par\end{centering}
\caption{In-plane MR ratio as a function of normalized nonmagnetic resistivity
$\rho_{0}/\rho_{{\rm m}}$, where $\rho_{0}=\sigma_{D}^{-1}$ and
$\rho_{{\rm m}}=(\sigma_{D}^{{\rm (m)}})^{-1}$. Inset shows the resistivity
$\rho_{xx}$ as a function of $\alpha$ for $\rho_{0}/\rho_{{\rm m}}=10$. }
\label{fig:MR ratio}
\end{figure}

The magnitude of the in-plane magnetoresistance can
be expressed in terms of the MR ratio
\begin{equation}
\mathrm{MR}=\frac{\rho_{xx}(\alpha=0)-\rho_{xx}(\alpha=\pi/2)}{\rho_{xx}(\alpha=0)},\label{ratio}
\end{equation}
where $\rho_{xx}=\sigma_{xx}/(\sigma_{xx}^{2}+\sigma_{xy}^{2})$ is
the resistivity. For comparison with experiments, we assume disorder
with both nonmagnetic and magnetic impurities, Eqs.~(\ref{normal impurity})
and (\ref{Vm}). Figure~\ref{fig:MR ratio} is a plot of the in-plane
MR ratio as a function of the normalized nonmagnetic resistivity $\rho_{0}/\rho_{{\rm m}}$,
where $\rho_{0}=\sigma_{D}^{-1}$ and $\rho_{{\rm m}}=(\sigma_{D}^{{\rm (m)}})^{-1}$.
The MR ratio can beomce 42.3\% in the absence of non-magnetic scattering
$\rho_{0}=0$, but gradually decreases with increasing $\rho_{0}/\rho_{{\rm m}}$.
Ref.~\onlinecite{Banerjee15} reports a measured $\mathrm{MR}\sim3\%$
for a TI$|$YIG system, while Ref.~\onlinecite{Yasuda16} finds $\mathrm{MR}\sim13\%$
for TI$|$magnetically-doped-TI bilayers. In the inset of Fig.~\ref{fig:MR ratio},
we plot the resistivity $\rho_{xx}$ as a function of $\alpha$ for
$\rho_{0}/\rho_{{\rm m}}=10$, in good agreement with the measured
in-plane MR of TI$|$CoFeB bilayers \cite{Lv17}.

\section{Summary}

\label{Summary}

We model the magnetic-proximity-induced magnetoresistance in disordered
topological
$\vert$
ferromagnetic insulator bilayers. Assuming that an FI contact magnetizes
the TI-surface states, we derive analytical expressions for the electric
dc conductivity. We formulate electron transport by the Kubo linear
response theory including the ladder-vertex correction as well as
by the Boltzmann approach including the in-scattering term of the
collision integral. The induced exchange splitting generates an electric
resistance that depends on the normal component of the magnetization
direction for non-magnetic disorder. For in-plane magnetizations,
unlike for the magnetic Rashba 2D system, the in-plane MR then vanishes.
For out-of-plane magnetizations, we predict that the gap opening at
the Dirac point causes a large MR ratio. On the other hand, we do
find an in-plane MR and planar Hall effect in the presence of magnetic
impurities aligned to the FI magnetization that can be explained by
the spin-momentum locking for Dirac electrons. Our calculated results
agree with the limited number of experiments on the out-of-plane MR.
We explain the MR\ observed for in-plane magnetizations by magnetic
disorder, which thereby provides information on the interface morphology.
In terms of the functional dependence on magnetization direction,
our model predicts a mixture of AMR and SMR character. Our model calculation
might help the theoretical design of topological insulators for next-generation
spin-based information technologies.

\section{Acknowledgments}

The authors thanks Y. Araki, K. Nomura, Y. Ohminato, I. Matsuzaki,
D. Kurebayashi, Y. Shiomi, J. Inoue, and M. Titov for valuable discussions.
This work was supported by Grants-in-Aid for Scientific Research (Grant
Nos. 25247056, 25220910, 268063) from the JSPS, FOM (Stichting voor
Fundamenteel Onderzoek der Materie), the ICC-IMR, EU-FET Grant InSpin
612759, and DFG Priority Programme 1538 ``Spin-Caloric Transport\textquotedblright \ (BA
2954/2).


\appendix

\section{Current-induced spin polarization in the presence of magnetic disorder}

\label{Appendix}

Here we calculate the electric-field-driven nonequilibrium spin density
in the presence of magnetic impurities {[}Eq.~(\ref{Vm}){]} by the
Boltzmann theory. A nonzero steady-state charge current implies a
finite spin density that can easily be found by multiplying the charge
current by $-\hbar/(2ev_{F})$, yielding 
\begin{equation}
\langle\mathbf{s}\rangle=e\sum_{s}\int\frac{d^{2}\mathbf{q}}{(2\pi)^{2}}\delta\left(E_{F}-E_{qs}\right)\tau_{\mathrm{m}}(\mathbf{q})(\mathbf{v}_{qs}\cdot\mathbf{E})\mathbf{s}_{qs},\label{sxy}
\end{equation}
where $\mathbf{s}_{qs}=-\hbar/(2ev_{F})\hat{\mathbf{z}}\times\mathbf{v}_{qs}$.
In the presence of an electric field $\mathbf{E}=E_{x}\hat{\mathbf{x}}$
and for in-plane magnetizations, 
\begin{align}
\langle\mathbf{s}\rangle= & -\frac{\hbar}{2}\frac{2eE_{x}}{hv_{F}}\frac{E_{F}\tau_{\mathrm{m}}^{\mathrm{e}}}{\hbar}\frac{\sqrt{3}}{3}\Bigl[-2\left(2-\sqrt{3}\right)M_{x}M_{y}\hat{\mathbf{x}}\nonumber \\
 & +\left\{ 1-\left(2-\sqrt{3}\right)\left(M_{x}^{2}-M_{y}^{2}\right)\right\} \hat{\mathbf{y}}\Bigl].\label{spin accumulation}
\end{align}
In contrast to the current-induced spin polarization for the normal
disorder, Eq.~(\ref{spin accumulation}) is affected by an in-plane
magnetization through the exchange interaction in Eq.~(\ref{Vm}).
\end{document}